\documentclass[aps,prl,twocolumn,superscriptaddress,showpacs,floatfix]{revtex4-1}
\usepackage{graphicx}
\usepackage{dcolumn}
\usepackage{bm}
\usepackage{slantsc}%
\usepackage{subfigure}
\usepackage{amssymb}
\usepackage{MnSymbol}
\usepackage{fix-cm}
\usepackage{natbib}
\newcommand{\beq}{\begin{equation}}
\newcommand{\eeq}{\end{equation}}
\newcommand{\bea}{\begin{eqnarray}}
\newcommand{\eea}{\end{eqnarray}}

\def\mpt{$\left<p_{\rm t}\right>$}
\def\dndeta{${\rm d}N_{\rm ch}/{\rm d}\eta$}
\def\delTpt{$\langle \Delta p_{t,i}, \Delta p_{t,j} \rangle$}
\begin{document}

\title{ Color reconnection and event-by-event fluctuations of mean transverse momentum in proton-proton collisions at $\sqrt{s}$~=~7 TeV in PYTHIA6.4 }

\author{Qingjun Liu}
\affiliation{
Department of Mathematics and Physics, \\
Beijing Institute of Petro-chemical Technology, Beijing 102617, People's Republic of China
\vspace*{1ex}}
\author{Wei-Qin Zhao}
\affiliation{
Institute of High Energy Physics, Chinese Academy of Sciences,
 P. O. Box 918(4), Beijing 100049, People's Republic of China
\vspace*{1ex}}

\date{\today}

\begin{abstract}
Based on PYTHIA6.428, event-by-event fluctuations of the mean transverse momentum in 
proton-proton collisions at $\sqrt{s}$~=~7~TeV are studied using several measures of the fluctuations.
The study compares results of the fluctuations for events where color reconnection is turned on,
with those where color reconnection is turned off.
The comparison reveals that some of the fluctuation measures are much more sensitive to color reconnection than
others. Through the comparison it is demonstrated for the first time that for proton-proton collisions at the Large Hadron Collider,
these sensitive measures of the mean transverse momentum
fluctuations can serve as alternative probes to color
reconnection and color reconnection is a non-trivial source of the event-by-event fluctuations.
\end{abstract}

%
%
\maketitle
\textit{Introduction:---} Event-by-event fluctuations of the mean transverse momentum (\mpt)
in proton-proton collisions have attracted lots of
attention~\cite{Abelev:2014ckr,Heckel:2015swa,Aduszkiewicz:2015jna,Adler:2003xq}. On one hand, it is believed that 
proton-proton collision may be used as a model-independent baseline
system for studying non-trivial correlations and fluctuations in heavy-ion collisions.
The non-trivial fluctuations or correlations in heavy-ion collisions would
display as a modification of the pattern for the fluctuations or correlations
in proton-proton collisions. Indeed, the modification was observed in collisions of Pb - Pb at $\sqrt{s_{NN}}=2.76$ TeV at the Large Hadron Collider(LHC)\cite{Abelev:2014ckr,Heckel:2015swa} and upto now satisfactory explanation for it is pending.
On the other hand, the measurement of the \mpt~fluctuations may aid the study of physical mechanisms for 
multi-particle production in proton-proton collisions. In Refs.~\cite{Abelev:2014ckr,Heckel:2015swa},
PYTHIA model\cite{Sjostrand:2006za,Skands:2010ak,Sjostrand:2007gs} predictions for 
\mpt~fluctuations in proton-proton collisions
were compared with data measured at the LHC. The study concluded that in proton-proton
collisions at the LHC, a \mpt~fluctuation measure~\cite{Voloshin:1999yf} well-known in the community of heavy-ion collisions~\cite{Stodolsky:1995tfm,Shuryak:1997yj,Stephanov:1998dy,Stephanov:1999zu,Basu:2016ibk,Tannenbaum:2005we}
is of poor sensitivity to a physical mechanism for
parton production called color reconnection(CR)~\cite{Sjostrand:2006za,Skands:2007zg,Sjostrand:2013cya,Christiansen:2015yqa}
The study also implies that in proton-proton at the LHC, CR contributes little to \mpt~fluctuations measured with
that particular measure. However,
several other measures~\cite{Gazdzicki:1992ri,Ray:2002md,Adams:2003uw,Trainor:2015swa,Adcox:2002pa,Adler:2003xq,Adamova:2003pz,Adamova:2008sx,Gorenstein:2011vq,Gazdzicki:2013ana} which have been extensively used for studying \mpt~fluctuations in proton-proton and heavy-ion collisions at the SPS, RHIC and LHC energy regime~\cite{Appelshauser:1999ft,Adamova:2003pz,Anticic:2003fd,Anticic:2008aa,Liu2003184,Liu:2008pv,Adcox:2002pa,Adler:2003xq,Ferreiro:2003dw,Adamova:2008sx,Gorenstein:2011vq,Gazdzicki:2013ana,Anticic:2015fla,Aduszkiewicz:2015jna}, have not been applied for
studying both CR and its effect on the fluctuations in proton-proton collisions yet.

Happening at the partonic stage of particle production just before hadronization in collisions with several multi-parton
inter-actions (MPI)\cite{Sjostrand:2006za}, CR
joins partons of low and high transverse momentum according to a calculated probability to connect partons\cite{Corke:2010yf,Sjostrand:2004pf}.
Since it was first studied in the
context of rearrangements of partons\cite{Gustafson:1988fs}, CR has been shown later strongly suppressed at the perturbative level\cite{Fadin:1993kt}.
Data measured at LHC\cite{Ortiz:2013yxa,Abelev:2013bla,Adam:2015gda} has begun
to witness the importance of CR in proton-proton, proton-nucleus and
nucleus-nucleus collisions.
However, the issue of CR in hadron collisions is extremely challenging\cite{Sjostrand:2006za},
let alone in heavy-ion collisions where large amout of MPI may occur.
Further theoretical and experimental studies about CR are necessary, including searching for sensitive observables and investigating CR-induced effects on the observables~\cite{Christiansen:2015cpa}.

In this letter we demonstrate for the first time that for proton-proton
collisions at the LHC, several measures of the event-by-event \mpt~fluctuations are quite sensitive to CR thus can serve
as alternative probes to CR, and applying these measures CR is demonstrated to be a non-trivial source of the event-by-event \mpt~fluctuations.

\textit{Analysis, Results and Discussions:---}All results presented in this paper are from analyzing proton-proton collisions at $\sqrt{s}$=7 TeV simulated by using Monte Carlo event generator PYTHIA6.428.
Used for the simulations are two tunes of PYTHIA6.428, Perugia 2011 Default where CR is switched on and Perugia 2011 NOCR where CR
is switched off. In the analysis, we use the charged-particles that
are in the kinematic range $|\eta|<0.8$ and $0.15<p_{\rm t}<2$~GeV/$c$, where $\eta$ and $p_{t}$ symbolize
the pseudo-rapidity and the transverse momentum of a charged particle, respectively. In this letter, event multiplicity refers to the number of charged particles in that kinematic range. Utilized in the analysis are $1000\times10^{6}$ and $1024\times10^{6}$ events generated by using Perugia 2011 Default and Perugia 2011 NOCR, respectively.

We first present results for event-by-event \mpt~fluctuation measure $C=\langle \Delta p_{{\rm t},i}, \Delta p_{{\rm t},j} \rangle$. Usually called transverse momentum correlator, through $C = \sigma_{\rm p_{t},dyn}^2$ it is directly related to the
dynamical component $\sigma_{\rm p_{t}, dyn}$~\cite{Voloshin:1999yf,Ray:2002md} of the \mpt~fluctuations. The measure has been used to study event-by-event \mpt~fluctuations in proton-proton and nucleus-nucleus collisions at the LHC, RHIC and SPS energies~\cite{Gavin:2003cb,Voloshin:2003ud,Abelev:2014ckr,Adams:2005ka,Adamova:2008sx,Liu:2008pv,Broniowski:2009fm,Bozek:2012fw,Adamczyk:2013up}.

$C_{m}$, being $C$ for events classified according to the event multiplicity, is defined as the mean of covariances of all pairs of particles $i$ and $j$ in the same event
with respect to the inclusive $\hat{p_{\rm t}}$ in event class ${\it m}$.
In order for the correlator $C_{m}$ not to be influenced by multiplicity fluctuations, each event class is built to have a unique event multiplicity.
This means that all events in a specific event class are of the same number of charged particles.
According to Ref. \cite{Voloshin:1999yf,Adams:2005ka,Abelev:2014ckr}, $C_{m}$ is formulated as
\begin{eqnarray}
C_m= (\sum\limits_{k=1}^{ \epsilon}{ \sum\limits_{i=1}^{n_{k} - 1}
{
\sum\limits_{j=i+1}^{n_{k}}
{
{\left(
{p_{t,i}-
\hat{p}_{t}
}
\right)}
{\left(
{p_{t,j}-
\hat{p}_{t}
 }
\right)}
}
}}/(\sum\limits_{k=1}^{\epsilon} {N_{k}^{\rm pairs}}).~~
\label{eq:correlator}
\end{eqnarray}
In Eq.~(\ref{eq:correlator}), $\epsilon$ is the number of events in event class $m$,
$n_{k}$ and $N_{k}^{\rm pairs} = 0.5 \cdot n_{k} \cdot (n_{k}-1)$ are the number of charged particles and charged particle pairs
in the $k^{th}$ event of event class $m$, and $\hat{p}_{t}$ is the average
transverse momentum of all particles in all events of class $m$. It is calculated according to 
\begin{equation}
  \hat{p}_{t}=(\sum\limits_{k=1}^{\epsilon}\sum\limits_{i=1}^{n_{k}} p_{{t},i})/(\sum\limits_{k=1}^{\epsilon} n_{k})=(\sum\limits_{k=1}^{\epsilon} n_k \cdot \left\langle {p_t} \right\rangle _k)/(\sum\limits_{k=1}^{\epsilon} n_{k}),
\label{eq:pt_hat}
\end{equation}
where $\left\langle {p_t} \right\rangle _k=\left(\sum_{i=1}^{n_{k}}
{p_{t,i}}\right) /n_k$ is the event-wise average transverse momentum \mpt~for the $k^{th}$ event and $p_{t,i}$ is the transverse momentum of the $i^{th}$ particle in that event.
By construction, $C_m$ vanishes in the case of uncorrelated particle emission, when only statistical fluctuations are present.
More details about the measure may be found in Ref.~\cite{Voloshin:1999yf,Abelev:2014ckr,Adams:2005ka}.

\begin{figure}[th]
\centering
\resizebox{\linewidth}{!}{
\includegraphics[scale=0.80]{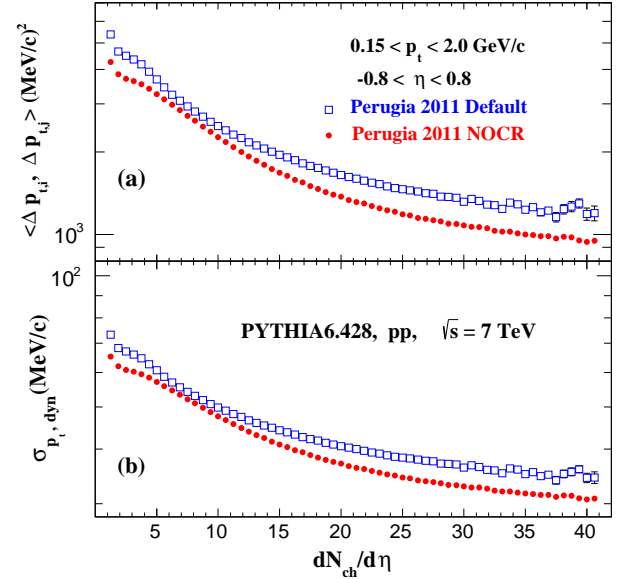}}
\caption{(Color online)
\delTpt~and $\sigma_{p_{t},dyn}$ as a function of charged
particle density \dndeta~in proton-proton collisions at $\sqrt{s}$~=~TeV in PYTHIA6.428.
\label{fig1} }
\end{figure}
In Fig.~\ref{fig1}(a), transverse momentum correlator \delTpt~is shown as a function of \dndeta.
It can be seen from
of Fig.~\ref{fig1}(a) that due to CR,
transverse momentum correlations are
enhanced non-trivially. The enhancement is more evident in events of greater multiplicity because in these events there are more multi-parton hard scatterings and thus more color reconnections that induce parton correlations through a large transverse boost\cite{Sjostrand:2007gs,Ortiz:2013yxa}. When these correlated partons hadronize, additional transverse momentum correlations among final state charged particles appear. That boost not only enlarges $\hat{p}_t$ but also $\sigma_{p_t}$. The errors shown in Fig.~\ref{fig1} are statistical and
are estimated by using a sub-event method. All the events falling 
in each of the bins defined according to $N_{ch}$, were divided into $M$ sub-sets with each set containing the same number of events.
We then analyzed each sub-set of the events. The mean of the results from the
$M$ sub-sets is the calculated result for all the events in
that bin, and its statistical error is calculated as the standard deviation
from the $M$ sub-sets divided by $\sqrt{M}$. In this paper we set $M$ to be 12. The errors shown in other figures of this paper were also obtained in this way. 
That \delTpt~equals to $\sigma_{p_{t},dyn}^2$~\cite{Voloshin:1999yf,Adams:2005ka,Adamova:2008sx} allows us to calculate $\sigma_{p_{t},dyn}$~as a function of \dndeta.
The calculated results are shown in Fig.~\ref{fig1}(b). From
Fig.~\ref{fig1}(b) one may tell that CR contributes
to $\sigma_{p_{t}, dyn}$ substantially. Therefore CR is a source
of the dynamical \mpt~fluctuations and it can be probed by \mpt~fluctuation measure $\sigma_{p_{t},dyn}$ or two particle transverse momentum correlator \delTpt~in proton-proton collisions at LHC.

\begin{figure}[th]
\resizebox{\linewidth}{!}{
\includegraphics[scale=0.80]{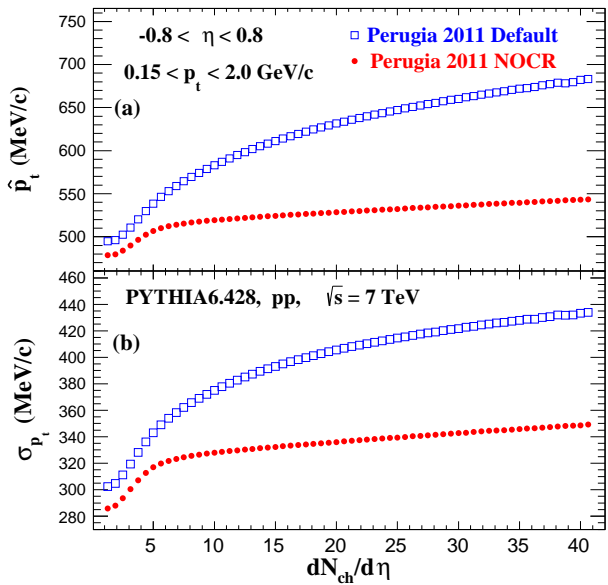}}
\caption{(Color online)
The mean $\hat{p}_{t}$ and the standard deviation $\sigma_{p_{t}}$ of the
 charged particle inclusive transverse momentum distributions, as a function
of charged particle density \dndeta~in proton-proton collisions at $\sqrt{s}$~=~7~TeV in PYTHIA6.428.
\label{fig2} }
\end{figure}
That CR has no big effect on
$\sqrt{\langle \Delta p_{t,i}, \Delta p_{t,j}\rangle}/\hat{p}_{t}$~\cite{Abelev:2014ckr,Heckel:2015swa} is understandable and allows us to learn more about $\sigma_{p_{t}, dyn}=\sqrt{C}=\sqrt{\langle \Delta p_{t,i}, \Delta p_{t,j} \rangle}$ as a measure for CR. 
To that end, we have also repeated the calculation of $\hat{p}_{t}$ and the result is displayed in Fig.\ref{fig2}(a).
As shown in Fig.~\ref{fig1}(b) and Fig.~\ref{fig2}(a), CR causes
both $\sqrt{\langle \Delta p_{t,i}, \Delta p_{t,j} \rangle}$ and $\hat{p}_{t}$~to increase, which is more evident in events of greater $N_{ch}$.
Thus one may note from a derivative of $\sqrt{\langle \Delta p_{t,i}, \Delta p_{t,j}\rangle}/\hat{p}_{t}$ that when CR increases both $\sqrt{C}$ and $\hat{p}_{t}$ at the rate of
\bea
\label{eq3}
\frac{\sqrt{C}|_{Default}- \sqrt{C}|_{NOCR}}{\hat{p}_{t}|_{Default}-\hat{p}_{t}|_{NOCR}}=\sqrt{C}/\hat{p}_{t},
\eea
then the ratio $\sqrt{\langle \Delta p_{t,i}, \Delta p_{t,j} \rangle}/\hat{p_{t}}$ does not vary appreciably with CR. Because $\sqrt{C}/\hat{p}_{t}$ is quite less than one\cite{Abelev:2014ckr} as one can also see by comparing Fig.~\ref{fig1}(b)
with Fig.~\ref{fig2}(a) for events of high multiplicity, Eq.~(\ref{eq3}) reveals that $\hat{p}_{t}$ is much more
sensitive to CR than $\sigma_{p_{t}, dyn}=\sqrt{C}$ is. However, one should be
reminded that $\sigma_{p_{t}, dyn}$ reflects the effect of CR on two-particle transverse momentum correlations while $\hat{p}_{t}$ does not, hence together with$\langle \Delta p_{t,i}, \Delta p_{t,j} \rangle$, it may allow one to study CR in another perspective.

The measure $\sqrt{\langle \Delta p_{t,i}, \Delta p_{t,j} \rangle}/\hat{p_{t}}$ for the event-by-event \mpt~fluctuations relates to the specific heat of the thermalized nuclear matter produced in high energy central collisions of
heavy nuclei~\cite{Stodolsky:1995tfm,Shuryak:1997yj,Stephanov:1998dy,Stephanov:1999zu,Basu:2016ibk,Tannenbaum:2005we}. According to Ref.~\cite{Abelev:2014ckr}, it is currently not a favored measure for studying CR in proton-proton at the LHC.
However, the standard deviation $\sigma_{p_t}$ of the charged particle inclusive
transverse momentum distributions may be a promising one instead, as can be
seen in Fig.~\ref{fig2}(b).
While $\hat{p_{t}}$~as a function of $N_{ch}$ helps validate the importance of CR\cite{Abelev:2014ckr,Skands:2010ak} as may also be seen in Fig.~\ref{fig2}(a),
Fig.\ref{fig1}(a) and Fig.~\ref{fig2}(b) also demonstrate respectively that
\delTpt~and $\sigma_{p_t}$ can be
useful observables constraining theoretical models of CR in
high energy proton-proton collisions at the LHC. Therefore we strongly recommend
that not only $\hat{p}_{t}$, but also both $\sigma_{p_t}$ and \delTpt~for charged particles in the kinematic range $|\eta|<0.8$ and $0.15<p_{\rm t}<2$~GeV/$c$ be measured all together in proton-proton collisions at the LHC.

\begin{figure}[hb]
\centering
\resizebox{\linewidth}{!}{
\includegraphics[scale=0.80]{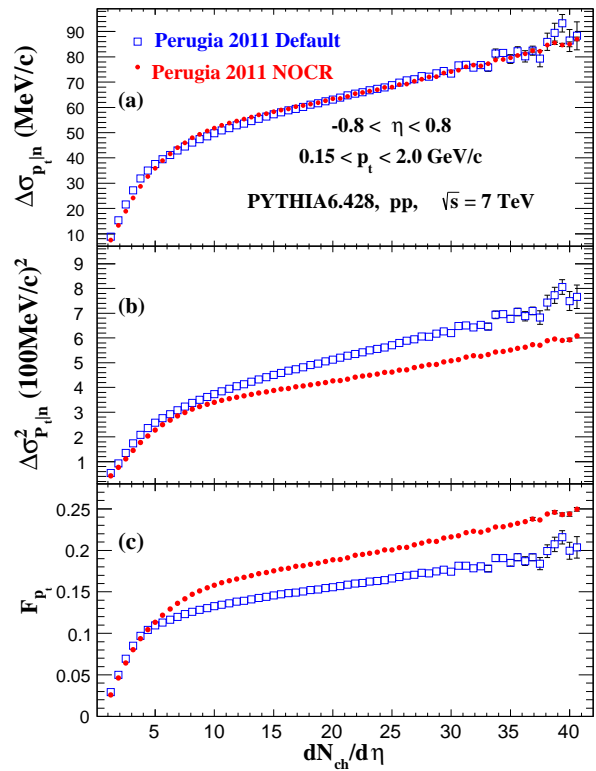}
}
\caption{(Color online)
$\Delta \sigma_{P_t|n}$, $\Delta \sigma^2_{P_t|n}$,
and $F_{p_t}$, as a function of charged particle density \dndeta~in
proton-proton collisions at $\sqrt{s}$~=~7~TeV in PYTHIA6.428.
\label{fig3} }
\end{figure}
We furthermore explore the effect of CR on \mpt~fluctuations by using other measures of the fluctuations.
Those measures include the $\Phi_{p_t}$\cite{Gazdzicki:1992ri}, $\Delta \sigma_{p_t:n}$ or $\Delta \sigma_{p_t}$\cite{Ray:2002md,Adams:2003uw}(now named $\Delta \sigma_{P_t|n}$\cite{Trainor:2015swa}), $\Delta \sigma^2_{P_t|n}$ and $\bar{B}$~\cite{Trainor:2015swa}, $F_{p_t}$\cite{Adcox:2002pa,Adler:2003xq}, $\Sigma_{p_t}$\cite{Adamova:2003pz,Adamova:2008sx},$\Delta[P_{t},n]$ and $\Sigma[P_{t},n]$\cite{Gorenstein:2011vq,Gazdzicki:2013ana}, where $n$ and $P_t$ represent the
event multiplicity and the total transverse momentum of charged particles in an event.
One may note that $n_{k}=N_{ch}=n$ and $P_{t}=N_{ch}$\mpt.
According to Refs.~\cite{Voloshin:1999yf,Ray:2002md,Adamova:2003pz,Trainor:2015swa,Gazdzicki:2013ana}, for a sample
of events with event multiplicity $n$, we formulated the interrelations among the \mpt~fluctuation measures as 
\bea
\nonumber \Delta \sigma^2_{P_t|n} &=&
(n-1) C = \bar{B}/n
\\
&=& 2\sigma_{p_t} \Delta \sigma_{P_t|n} \approx 2\sigma_{p_t}\Phi_{p_t}
\approx 2\sigma_{p_t}^{2} F_{p_t}~.
\label{eq4}
\eea
\bea
\label{eq5}
\sqrt{\Sigma[P_{t},n]} &=& \sqrt{\Delta[P_{t},n]}=\Phi_{p_t}/\sigma_{p_t}+1
\approx F_{p_t}+1~.
\eea
\bea
\label{eq6}
\Sigma_{p_t} = \sqrt{(n-1)/n} \sqrt{C}/\hat{p_t}. 
\eea
Shown in Fig.~\ref{fig3} are $\Delta \sigma_{P_t|n}$, $\Delta \sigma^2_{P_t|n}$
and $F_{p_t}$, which were calculated applying both Eq.~(\ref{eq4}) and the results
already obtained for $C$ and $\sigma_{p_t}$.

\begin{figure}[th]
\centering
\resizebox{\linewidth}{!}{
\includegraphics[scale=0.80]{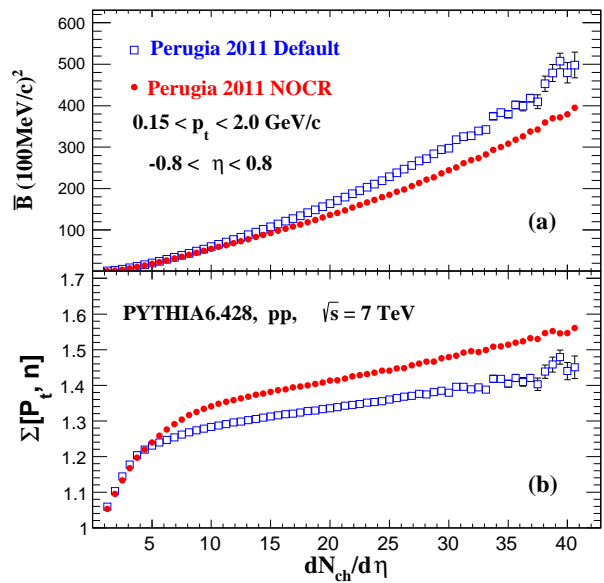}
}
\caption{(Color online)
$\bar{B}$ and $\Sigma[P_t,~n]$
 as a function of charged particle density \dndeta~in
proton-proton collisions at $\sqrt{s}$ = 7 TeV in PYTHIA6.428.
\label{fig4} }
\end{figure}

It can be seen from
Fig.~\ref{fig3} that $\Delta \sigma_{P_t|n}$ is of no sensitivity to CR
in the whole explored range of event multiplicity while $\Delta \sigma^2_{P_t|n}$ and $F_{p_t}$ are quite sensitive to CR in high
multiplicity events. Eq.~(\ref{eq6}) reveals that $\Sigma_{p_t}$ is no more sensitive
to CR than $\sqrt{\langle \Delta p_{t,i}, \Delta p_{t,j} \rangle}/\hat{p_{t}}$ is.
Henceforth $\Sigma_{p_t}$ and $\Delta \sigma_{P_t|n}$, together with $\Phi_{p_t}$ being approximately equivalent
to $\Delta \sigma_{P_t|n}$ as pointed out by Eq.~(\ref{eq4}), are not suitable for studying CR
, and \mpt~fluctuations studied by using these three
measures are not influenced by CR.  
Because $\Delta \sigma_{P_t|n}$ does not vary with CR and is defined as $\Delta \sigma^2_{P_t|n}/2\sigma_{p_{t}}$
as shown in Eq.~(\ref{eq4}), one may get
\bea
\label{eq7}
\frac{\Delta \sigma^2_{P_t|n}|_{Default}-\Delta \sigma^2_{P_t|n}|_{NOCR}}{2(\sigma_{p_{t}}|_{Default}-\sigma_{p_{t}}|_{NOCR})}= \Delta \sigma_{P_t|n}.
\eea
Since $\Delta \sigma_{P_t|n}$ becomes much greater than $1$ in events of greater event multiplicity as shown in Fig.~\ref{fig3},
Eq.~(\ref{eq7}) tells that $\Delta \sigma^2_{P_t|n}$ is a lot more sensitive to CR than $\sigma_{p_{t}}$ is and the sensitivity increases with the increase of $N_{ch}$.
Furthermore from Eq.~(\ref{eq4}) and Eq.~(\ref{eq5}), one may get $\Sigma[P_{t},n]=\Delta[P_{t},n]=(1+F_{p_t})^{2}$ and $\bar{B}=n\Delta \sigma^2_{P_t|n}$, which reveal that $\Sigma[P_{t},n]$ and $\Delta[P_{t},n]$ are even more sensitive to CR
than $F_{p_t}$ is, and $\bar{B}$ has a better sensitivity to CR than $\Delta \sigma^2_{P_t|n}$ does. That both $\bar{B}$ and $\Sigma[P_{t},n]=\Delta[P_{t},n]$ are quite sensitive to CR may be clearly seen in Fig.~\ref{fig4}. Additionally from Eq.~(\ref{eq4}) one gets $C=\Delta \sigma^2_{P_t|n}/(n-1)$, which indicates that with the increasing of event multiplicity $C$ becomes being less sensitive to CR than $\Delta \sigma^2_{P_t|n}$ is.    
Furthermore, from Eq.~(\ref{eq4}), one may get $F_{p_t}\approx \Delta \sigma_{P_t|n}/\sigma_{p_{t}}$ and $\Delta \sigma^2_{P_t|n}=2\sigma_{p_{t}}\Delta \sigma_{P_t|n}$.
Taking into considerations again the fact that $\Delta \sigma_{P_t|n}$ does not vary with CR, these two expressions reveal that the sensitivity of both $F_{p_t}$ and $\Delta \sigma^2_{P_t|n}$ to 
CR is not all but partly attributed to $\sigma_{p_{t}}$ being sensitive to CR, which
has already been demonstrated by Fig.~\ref{fig2}(b). This also holds true for $\bar{B}$ and $\Sigma[P_{t},n]=\Delta[P_{t},n]$ because $\bar{B}=2n\sigma_{p_{t}}\Delta \sigma_{P_t|n}$ and $\sqrt{\Sigma[P_{t},n]}=\sqrt{\Delta[P_{t},n]}\approx \Delta \sigma_{P_t|n}/\sigma_{p_{t}}+1$ according to Eq.~(\ref{eq4}) and Eq.~(\ref{eq5}).   
Therefore $F_{p_t}$, both $\Sigma[P_{t},n]$ and $\Delta[P_{t},n]$, $\bar{B}$, 
$\Delta \sigma^2_{P_t|n}$ and $C$ are preferred for studying CR
through measuring \mpt~fluctuations in proton-proton collisions at the LHC and
CR adds significantly to the \mpt~fluctuations
in the collisions of high event multiplicity when the fluctuations are analyzed using these measures.

\textit{Summary:---} We present an analysis of event-by-event fluctuations of the mean transverse momentum for PYTHIA6.428-simulated proton-proton collisions at $\sqrt{s}$~=~7 TeV by
using various measures of the event-by-event fluctuations. 
Two tunes of PYTHIA6.428, Perugia 2011 Default and Perugia 2011 NOCR, were used
for simulating the collisions.
The results from the Perugia 2011 Default where color reconnection is turned on 
have been compared with those from Perugia 2011 NOCR where color
reconnection is turned off. The comparison has clearly indicated
that while measures $F_{p_t}$, both $\Sigma[P_{t},n]$ and $\Delta[P_{t},n]$,
$\Delta \sigma^2_{P_t|n}$, $\bar{B}$ and $C$ are quite sensitive to color reconnection in collisions of high event multiplicity, $\Sigma_{p_t}$, $\Delta\sigma_{P_{t}|n}$ and $\Phi_{p_t}$ are not at all in our analysis.
Therefore these sensitive measures are promising probes to color reconnection
and
color reconnection is a non-trivial source of the event-by-event fluctuations
of the mean transverse momentum in proton-proton and heavy-ion collisions at the LHC when the event-by-event fluctuations are analyzed with these sensitive measures.
Furthermore $\sigma_{p_t}$ has been shown to be quite sensitive to color reconnection, hence it is strongly recommended that $\sigma_{p_t}$ be measured that
may also help constrain color reconnection model in proton-proton collisions at the LHC.

\textit{Acknowledgements:} This work was supported in part by Beijing Natural
Science Foundation under Project No. 1132017, National Natural Science Foundation of China, and the Scientific Research Foundation for
the Returned Overseas Chinese Scholars, State Education Ministry. Computing support is partly provided by the Supercomputing 
Center, Computer Network Information Center, Chinese Academy of Sciences. 

\nocite{*}
\bibliographystyle{apsrev4-1}
\bibliography{crebe}
\end{document}